\documentclass{article}

\PassOptionsToPackage{numbers,compress,sort}{natbib}
\usepackage[final,main]{neurips_2026}
\usepackage{amsmath}
\usepackage{amssymb}
\usepackage{graphicx}
\graphicspath{{figures/}}
\usepackage{booktabs}
\usepackage{array}
\usepackage{multirow}
\usepackage{xcolor}
\usepackage[colorlinks=true,linkcolor=blue,citecolor=blue,urlcolor=blue]{hyperref}
\usepackage{algorithm}
\usepackage{algorithmic}
\usepackage{subcaption}
\usepackage{microtype}
\usepackage{bm}
\usepackage{nicefrac}
\usepackage{placeins}

\newcommand{\ours}{\textsc{NeuroHorizon}}
\newcommand{\poyoplus}{\textsc{POYO+}}
\newcommand{\fpbps}{\text{fp-bps}}

\title{NeuroHorizon: Long-Horizon Forecasting of Neural Population Activity with Hierarchical Memory}

\author{%
  Peng Jiang\textsuperscript{1,2,3} \\
  \texttt{jiang-p21@mails.tsinghua.edu.cn}
  \And
  Xiaoxuan Jia\textsuperscript{1,2,3}\thanks{Corresponding author.} \\
  \texttt{jiaxx@mail.tsinghua.edu.cn}
}

\hypersetup{
  pdftitle={NeuroHorizon: Long-Horizon Forecasting of Neural Population Activity with Hierarchical Memory},
  pdfauthor={Peng Jiang; Xiaoxuan Jia},
  pdfsubject={NeurIPS 2026 Main Track camera-ready},
  pdfkeywords={neural population forecasting, autoregressive rollout, hierarchical memory}
}

\begin{document}

\maketitle
\footnotetext[1]{Center for Life Sciences and Artificial Intelligence, School of Life Sciences, Tsinghua University, Beijing, China.}
\footnotetext[2]{IDG/McGovern Institute for Brain Research, Tsinghua University, Beijing, China.}
\footnotetext[3]{Tsinghua-Peking Center for Life Sciences, Tsinghua University, Beijing, China.}

\begingroup
\renewcommand{\thefootnote}{}
\footnotetext[0]{Code: \url{https://github.com/Jp-17/NeuroHorizon}}
\endgroup

\begin{abstract}
Long-horizon forecasting of neural population activity tests whether a model has captured dynamics that remain predictive well beyond the observed history.
Prediction quality typically declines as the horizon grows, yet neither short-horizon accuracy nor teacher-forced evaluation, which supplies the true activity preceding each predicted step, reveals how well models sustain predictions from an initial history alone.
We compare direct multi-step prediction and autoregressive rollout on spiking recordings across horizons and introduce \ours{}, which pairs event-based spike encoding with a causal decoder for population firing rates.
Its hierarchical prediction memory keeps recent predictions at full temporal resolution and compresses earlier ones into segment summaries, giving the decoder both local detail and longer-range context.
Under teacher-forced training this memory holds ground-truth activity, whereas during rollout it holds the model's own predictions.
To reduce this mismatch, we train with scheduled sampling, progressively exposing the decoder to its own predictions.
From 500\,ms of macaque motor-cortex activity, \ours{} achieves the highest forward-prediction score among the evaluated autonomous models at all four horizons from 100 to 1000\,ms, and its per-bin predictions remain above a mean-firing-rate baseline across nearly the full 1\,s window.
Complementary experiments on macaque sensorimotor and mouse multi-region recordings show that performance generally increases with model and data size, alongside improved prediction after cross-subject pretraining and fine-tuning.
Together, these results clarify how forecasting performance changes across horizons and offer an effective approach to long-horizon forecasting of neural population activity.
\end{abstract}

\section{Introduction}
\label{sec:introduction}

Predicting future neural population activity from past spikes provides a quantitative test of learned neural dynamics~\citep{pandarinath2018inferring,pei2021neural}.
Long-horizon prediction extends this test by asking how long those dynamics remain informative without new observations, with potential relevance to anticipatory neural control~\citep{lu2025benchmarking}.
Because predictive performance typically declines as the forecast horizon increases, models should be evaluated both on their overall performance and on how well they sustain predictions over time.

Autonomous forecasting can use direct multi-step prediction, defined here as jointly predicting the future window, or autoregressive rollout, which advances through generated outputs or latent states~\citep{taieb2012review}.
In both cases, predictions depend only on the initial neural history.
Existing neural population models include LFADS, masked transformers such as IBL-MTM and NDT2, and autoregressive event generators such as Neuroformer~\citep{pandarinath2018inferring,ibl2024mtm,ye2023ndt2,ye2023neuroformer}.
Forward prediction has been evaluated over short fixed windows, including 200\,ms in NLB and IBL-MTM~\citep{pei2021neural,ibl2024mtm}.
Teacher-forced evaluation instead supplies true future prefixes, so its scores do not measure forecasting performance from the initial history alone~\citep{bengio2015scheduled}.
The resulting question is how these forecasting strategies compare as the horizon increases when no further neural observations are available.

Calcium-imaging studies have already examined forecasting over longer intervals.
POCO and ZAPBench evaluate future neural activity across extended windows~\citep{duan2025poco,lueckmann2025zapbench}, and Lu et al.\ report informative forecasts of mouse cortical activity up to approximately 1.5\,s~\citep{lu2025benchmarking}.
However, calcium fluorescence reflects temporally filtered spiking activity, with calcium and indicator kinetics contributing persistence to the measured signal~\citep{chen2013ultrasensitive,deneux2016accurate}.
These findings do not directly establish how far into the future neural activity can be predicted from spike recordings alone.

To assess long-horizon forecasting from spike recordings, we compare direct multi-step prediction and autoregressive rollout across sub-second to second-scale horizons using only an initial observation window.
Direct prediction avoids feeding earlier prediction errors back into subsequent steps, but must infer the entire future trajectory from the observed history.
Autoregressive rollout instead constructs this trajectory sequentially, allowing later predictions to depend on earlier generated activity while also permitting errors to propagate through that feedback~\citep{taieb2012review}.
When a model is trained only on true prefixes, replacing them with predicted prefixes during rollout introduces an additional training--inference mismatch, commonly termed exposure bias~\citep{bengio2015scheduled,ranzato2016sequence}.
This motivates an autoregressive approach that combines a structured representation of the predicted trajectory with training that exposes the model to its own predictions.

We introduce \ours{}, an encoder--decoder model for long-horizon forecasting of neural population activity.
A \poyoplus{}-style event encoder represents the observed spike history, and a causal decoder predicts future population firing rates one bin at a time~\citep{azabou2025poyo}.
The decoder uses a hierarchical prediction memory that retains recent predicted bins at full temporal resolution and summarizes the preceding trajectory at a coarser resolution, drawing on the use of compressed memory for long-range sequence modeling~\citep{rae2020compressive}.
This organization preserves local temporal detail while providing compact access to longer-range context.
During autonomous rollout, the memory is constructed from model predictions, while the observed history remains available through the encoder latents.
Scheduled sampling progressively increases the probability of feeding predictions rather than ground-truth counts into memory during training, exposing the decoder to the inputs it will encounter during rollout~\citep{bengio2015scheduled}.

We compare \ours{} with direct multi-step predictors and rollout-based models on the Perich--Miller dataset of macaque motor-cortex recordings~\citep{perich2025dandi}, using 500\,ms of history and four prediction horizons from 100 to 1000\,ms.
\ours{} achieves the highest mean forward-prediction score among the evaluated autonomous models at every horizon.
Beyond whole-window scores, per-bin evaluation shows predictions above a mean-firing-rate baseline across nearly the full 1\,s window.
In ablations, the combined memory yields higher mean rollout scores than either memory component alone, and scheduled sampling improves long-horizon rollout performance.
Complementary experiments examine model-size and data-size scaling on macaque sensorimotor and mouse multi-region recordings, and show improved cross-subject rollout prediction at 500\,ms after pretraining and fine-tuning.

\textbf{Our main contributions are summarized below.}
\begingroup
\setlength{\topsep}{3pt}
\setlength{\partopsep}{0pt}
\setlength{\itemsep}{3pt}
\setlength{\parsep}{0pt}
\begin{itemize}
    \item \textbf{Multi-horizon evaluation of spike-based forecasting.} We compare direct multi-step prediction and autoregressive rollout using the same initial history across models, evaluating whole-window performance and per-bin prediction decay while reporting teacher-forced evaluation separately.
    \item \textbf{Hierarchical memory for long-horizon forecasting.} We introduce \ours{}, combining prediction memory at two temporal resolutions with scheduled-sampling training to improve autonomous forecasting. The model achieves the highest mean scores at all four evaluated horizons in the comparison on the Perich--Miller dataset.
    \item \textbf{Scaling and cross-subject transfer.} We examine model-size and data-size scaling on macaque and mouse recordings, and observe improved 500\,ms rollout prediction after cross-subject pretraining and fine-tuning.
\end{itemize}
\endgroup

\section{Related Work}
\label{sec:related_work}

\textbf{Neural population modeling.}
Neural population models capture dependencies across neurons and time through different representations and learning objectives.
Point-process generalized linear models describe history-dependent spiking and coupling between neurons, while LFADS learns latent dynamics underlying population activity~\citep{truccolo2005point,pandarinath2018inferring}.
Transformer-based approaches such as NDT2 and IBL-MTM learn population representations through masked modeling across sessions~\citep{ye2023ndt2,ibl2024mtm}.
At the event level, Neuroformer generates neural activity autoregressively, including extended stimulus-conditioned simulations of events inferred from calcium imaging~\citep{ye2023neuroformer}.
Event-based representations also support behavioral decoding, as demonstrated by POYO and \poyoplus{} across heterogeneous neural populations~\citep{azabou2023unified,azabou2025poyo}.
These advances in neural representation learning also provide a basis for studying how well population activity can be predicted across different horizons.
Building on the \poyoplus{} encoder, we represent the observed spike history and use a causal decoder to predict future population firing rates.

\textbf{Neural activity forecasting and evaluation.}
Neural activity forecasting has been evaluated under varied observation conditions, prediction horizons, and performance measures.
For spike recordings, the Neural Latents Benchmark evaluates held-out future activity, XFADS reports 500\,ms forecasts from a 200\,ms initial trial segment on MC\_Maze, and SpikeProphecy benchmarks next-bin spike-count prediction from recorded history and analyzes its degradation during multi-step rollout~\citep{pei2021neural,dowling2024xfads,minnick2026spikeprophecy}.
For calcium recordings, POCO conditions neuron-specific forecasts on population activity, ZAPBench and Lu et al.\ benchmark whole-brain and probabilistic forecasting models, and OmniMouse includes a one-second forecasting task on calcium-derived activity~\citep{duan2025poco,lueckmann2025zapbench,lu2025benchmarking,willeke2026omnimouse}.
Building on these settings, we focus on spike recordings and ask how forecasting performance degrades as models predict further ahead without new observations, and whether this degradation differs between direct multi-step prediction and autoregressive rollout.

\textbf{Memory and exposure bias in autoregressive forecasting.}
Long-horizon autoregressive forecasting raises two problems, namely how to represent an expanding prediction history and how to train models to condition on their own outputs.
Transformer-XL represents long sequences by reusing hidden states across segments, while Compressive Transformer retains recent memories alongside compressed older representations~\citep{dai2019transformerxl,rae2020compressive}.
Exposure bias arises when a model trained on true prefixes must instead condition on its own outputs during inference, which can amplify errors over successive prediction steps~\citep{bengio2015scheduled,ranzato2016sequence}.
Scheduled sampling addresses this mismatch by introducing model predictions as inputs during training~\citep{bengio2015scheduled}.
We adapt these ideas in \ours{} by combining hierarchical prediction memory with scheduled sampling for neural population forecasting.

\section{Method}
\label{sec:method}

\begin{figure}[t]
    \centering
    \includegraphics[width=0.92\textwidth]{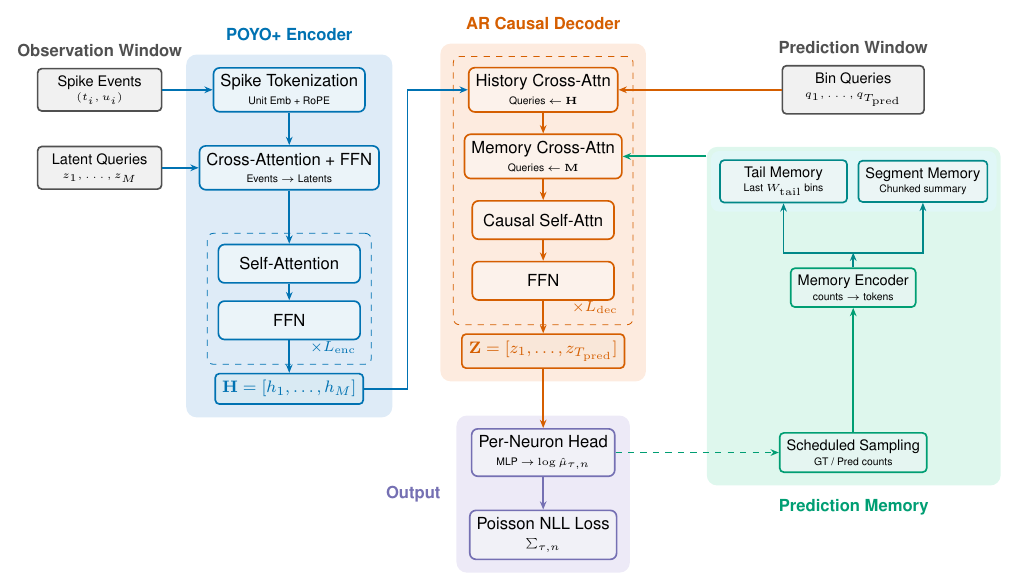}
    \caption{\textbf{Overview of the \ours{} architecture.} A \poyoplus{} event encoder maps observed spike events to history latents. A causal decoder predicts future firing rates bin by bin, attending to both history latents and a hierarchical prediction memory. During training, scheduled sampling chooses whether memory receives ground-truth counts or model predictions.}
    \label{fig:architecture}
\end{figure}

\ours{} encodes observed spike events into fixed history latents and predicts future population activity autoregressively (Figure~\ref{fig:architecture}).
Within this framework, a hierarchical memory represents the growing prediction prefix, while scheduled sampling exposes the decoder to predicted inputs during training.
Together, these components condition each prediction on the observed history and the preceding forecast.

\subsection{Forecasting Task and Objective}
\label{sec:problem}

We observe spike events within $[-T_{\mathrm{obs}}\Delta t,0)$ and forecast the next $T_{\mathrm{pred}}$ bins without further observations.
Here $T_{\mathrm{obs}}$ and $T_{\mathrm{pred}}$ specify window lengths in units of $\Delta t=20$\,ms, while the observed events retain their continuous timestamps.
Prediction targets are spike counts in these future bins.
Let $\bm y_\tau\in\mathbb N_0^N$ denote the counts for $N$ recorded neurons at prediction bin $\tau=1,\ldots,T_{\mathrm{pred}}$.
We use a conditional Poisson observation model $y_{\tau,n}\sim\operatorname{Poisson}(\mu_{\tau,n})$~\citep{truccolo2005point}, where $\mu_{\tau,n}=\lambda_{\tau,n}\Delta t$ for firing rate $\lambda_{\tau,n}$.
The decoder outputs the log expected count $\ell_{\tau,n}=\log\hat\mu_{\tau,n}$.

\textbf{Training objective.}
To fit the predicted means to observed spike counts, the training objective is the Poisson negative log-likelihood averaged over prediction bins and recorded neurons,
\begin{equation}
    \mathcal L_{\mathrm{train}}
    =\frac{1}{NT_{\mathrm{pred}}}
    \sum_{\tau=1}^{T_{\mathrm{pred}}}\sum_{n=1}^N
    \left[\hat\mu_{\tau,n}-y_{\tau,n}\log\hat\mu_{\tau,n}
    +\log(y_{\tau,n}!)\right].
    \label{eq:poisson_nll}
\end{equation}
The term $\log(y_{\tau,n}!)$ depends only on the observed counts and is omitted during optimization.
For minibatches with different population sizes, we average the loss over all valid neuron-bin entries, excluding padded neurons.

\subsection{Event Encoder and Causal Decoder}
\label{sec:architecture}

\textbf{Event encoder.}
To represent recordings with different neural populations in a common form, we encode spike timing and neuron identity into a fixed set of latent tokens.
Following \poyoplus{}~\citep{azabou2025poyo}, we represent observed spikes using learned neuron identity embeddings and continuous-time rotary positional encoding (RoPE).
Perceiver-style cross-attention~\citep{jaegle2021perceiver_original} and a feed-forward network (FFN), followed by $L_{\mathrm{enc}}$ self-attention--FFN blocks~\citep{vaswani2017attention}, map the events to history latents $\bm H\in\mathbb R^{M\times d}$.

\textbf{Causal decoder.}
The decoder conditions each future bin on both the observed activity and the preceding forecast through two cross-attention pathways.
History cross-attention provides a fixed reference to the observations, while memory cross-attention incorporates context accumulated over preceding prediction bins.
Causal self-attention then relates current and earlier bin representations without accessing future bins.
Each prediction bin starts from a learned query with RoPE evaluated at its center timestamp.
Each decoder layer updates its input states $\bm Q$ through history cross-attention, memory cross-attention, causal self-attention, and an FFN, with residual connections,
\begin{equation}
    \begin{aligned}
    \bm C &= \bm Q+\operatorname{HistCrossAttn}(\bm Q,\bm H),\\
    \bm V &= \bm C+\operatorname{MemCrossAttn}(\bm C,\{\bm M_\tau\}_\tau),\\
    \bm U &= \bm V+\operatorname{CausalSelfAttn}(\bm V),\\
    \bm Z &= \bm U+\operatorname{FFN}(\bm U).
    \end{aligned}
    \label{eq:decoder}
\end{equation}
To translate the shared population context encoded in each final bin state into neuron-specific predictions, a shared MLP combines that state with the target neuron's identity embedding and outputs its log expected spike count.

\subsection{Hierarchical Prediction Memory}
\label{sec:memory}

The prediction prefix contains both recent changes and activity patterns spanning multiple bins.
We represent these at two temporal resolutions so that the decoder can use local detail together with broader population context.
Let $\bm x_\tau$ denote the population vector written to memory after predicting bin $\tau$.
It contains ground-truth counts or predicted expected counts during training, and predicted expected counts during rollout.
The training feedback is selected by scheduled sampling (Section~\ref{sec:scheduled_sampling}).

\textbf{Memory encoder.}
The decoder needs a memory representation whose size does not depend on the number of recorded neurons.
We therefore combine each neuron's count with its identity embedding and pool the resulting features using learned summary queries.
The queries attend to neuron features as keys and values, producing a fixed number of tokens for each population vector.
The tail and segment encoders, $\phi_{\mathrm{tail}}$ and $\phi_{\mathrm{seg}}$, use this construction with independent parameters and produce $K_{\mathrm{tail}}$ tokens per tail bin and $K_{\mathrm{seg}}$ tokens per segment, respectively.

\textbf{Tail and segment memory.}
To preserve recent temporal detail, the tail keeps a separate representation for each of the most recent $W_{\mathrm{tail}}$ bins.
At prediction step $\tau$, these representations are concatenated as
\begin{equation}
    \bm M_\tau^{\mathrm{tail}}
    =\operatorname{Concat}_{u=\max(1,\tau-W_{\mathrm{tail}})}^{\tau-1}
    \phi_{\mathrm{tail}}(\bm x_u).
    \label{eq:tail_memory}
\end{equation}
To provide context beyond the recent window, the segment pathway summarizes the full available prefix using averages over consecutive groups of up to $S$ bins.
These summaries preserve a broader view of population activity without keeping a separate token group for every older bin.
Let $\bar{\bm x}_{j,\tau}$ denote the mean population vector of segment $j$ among the $J_\tau$ segments available before step $\tau$.
The segment tokens and the combined memory are
\begin{equation}
    \bm M_\tau^{\mathrm{seg}}
    =\operatorname{Concat}_{j=1}^{J_\tau}
    \phi_{\mathrm{seg}}(\bar{\bm x}_{j,\tau}),\qquad
    \bm M_\tau=[\bm M_\tau^{\mathrm{seg}};\bm M_\tau^{\mathrm{tail}}].
    \label{eq:segment_memory}
\end{equation}

\textbf{Memory regularization.}
Because generated feedback can be inaccurate, we train the decoder with perturbed or masked prediction memory.
During training, we perturb memory-encoder input features with noise and random zeroing.
We also apply block dropout to memory cross-attention, masking all tail and segment tokens for selected queries.
This trains the decoder to use observed history and causal bin representations when prediction memory is unavailable.

\subsection{Scheduled Sampling}
\label{sec:scheduled_sampling}

Training exclusively on ground-truth prefixes does not expose the decoder to the predicted inputs used during rollout.
Scheduled sampling~\citep{bengio2015scheduled} reduces this mismatch by selecting the memory source independently for each example and prediction bin.
At training epoch $e$, predicted feedback is selected with probability $\epsilon_e$,
\begin{equation}
    \bm x_\tau=
    \begin{cases}
        \bm y_\tau & \text{with probability }1-\epsilon_e,\\
        \operatorname{sg}(\hat{\bm\mu}_\tau) & \text{with probability }\epsilon_e.
    \end{cases}
    \label{eq:ss}
\end{equation}
The choice applies to the entire population vector.
The stop-gradient operator $\operatorname{sg}$ prevents later losses from propagating through the feedback values into earlier predictions, while the memory encoders remain trainable.
To introduce predicted feedback gradually, we increase its probability over $E_{\mathrm{ramp}}$ training epochs to a maximum $\epsilon_{\max}$,
\begin{equation}
    \epsilon_e=\epsilon_{\max}\min(e/E_{\mathrm{ramp}},1).
    \label{eq:ss_schedule}
\end{equation}
The probability increases linearly from zero and remains at $\epsilon_{\max}$ after the ramp.

\section{Experiments}
\label{sec:experiments}

\subsection{Datasets}
\label{sec:dataset}

\begin{table}[t]
    \centering
    \caption{\textbf{Dataset overview.} Neural population spiking datasets used in the main experiments. Session and subject counts refer to the recordings used in this study.}
    \label{tab:dataset_overview}
    \small
    \setlength{\tabcolsep}{3pt}
    \begin{tabular}{@{}>{\raggedright\arraybackslash}p{4.5cm} c c p{2.3cm} p{3.65cm}@{}}
        \toprule
        \textbf{Dataset} & \textbf{Sessions} & \textbf{Subjects} & \textbf{Region(s)} & \textbf{Task} \\
        \midrule
        Perich--Miller dataset~\citep{perich2025dandi} & 10 & 3 & M1 + PMd & \begin{tabular}[t]{@{}l@{}}Instructed-delay center-out\\reaching\end{tabular} \\
        O'Doherty--Sabes dataset~\citep{odoherty2017nonhuman} & 47 & 2 & M1 + S1 & \begin{tabular}[t]{@{}l@{}}Continuous hand-controlled\\2D cursor reaching\end{tabular} \\
        Allen Neuropixels dataset~\citep{siegle2021survey} & 32 & 32 & VIS, CA1, etc. & \begin{tabular}[t]{@{}l@{}}Passive natural movie\\viewing\end{tabular} \\
        \bottomrule
    \end{tabular}
\end{table}

We use three neural population datasets for the principal experiments (Table~\ref{tab:dataset_overview}).
The \textbf{Perich--Miller dataset}~\citep{perich2025dandi} supports our comparison across prediction horizons, with 10 sessions of structured center-out reaching from three macaques. We access these recordings through brainsets~\citep{azabou2023unified}.
The \textbf{O'Doherty--Sabes dataset}~\citep{odoherty2017nonhuman} comprises 47 sessions from two macaques performing continuous 2-D cursor reaching controlled by hand or finger movements, with recordings from M1 and S1.
We use 32 sessions from 32 mice in the \textbf{Allen Brain Observatory Visual Coding Neuropixels dataset}~\citep{siegle2021survey} during passive natural-movie viewing. The recordings span multiple brain regions, including visual cortical, thalamic, and hippocampal areas.
We use the O'Doherty--Sabes dataset and the Allen Neuropixels dataset to examine the effects of model size and training data size.
We additionally evaluate 500\,ms forecasting on MC\_Maze from the Neural Latents Benchmark (NLB)~\citep{pei2021neural}.

For comparisons on the Perich--Miller dataset, each window contains a 500\,ms observation period followed by a prediction horizon of 100, 240, 500, or 1000\,ms.
Training windows are sampled with randomized offsets within the training domains and shuffled each epoch.
Validation and test windows are sampled sequentially within their assigned trial intervals.
At each prediction horizon, all models share the same data splits, evaluation windows, null model, and metrics.
Dataset-specific splits are described in Appendix~\ref{app:data_splits}.

\subsection{Evaluation Modes and Metrics}
\label{sec:eval_modes}

We distinguish three prediction modes.
\textbf{Direct} forecasting predicts the entire future window from the observed history without feeding predictions back.
\textbf{Rollout (RO)} advances through generated predictions or an autonomously evolving latent state.
Direct forecasting and rollout are autonomous because they use only the initial neural history, without further observations.
\textbf{Teacher-forced (TF)} evaluation instead supplies the true activity preceding each prediction step.
We therefore report TF results separately from autonomous forecasting results.

The primary metric is forward-prediction bits-per-spike (\fpbps{})~\citep{pei2021neural}.
For each session, we compute
\begin{equation}
    \fpbps=\frac{\mathcal L_{\mathrm{null}}-\mathcal L_{\mathrm{model}}}
    {\log 2\cdot\sum_{w,\tau,n}y_{w,\tau,n}}.
    \label{eq:fpbps}
\end{equation}
Here $w$ indexes evaluation windows, $\tau$ prediction bins, and $n$ recorded neurons within the session.
$\mathcal L_{\mathrm{model}}$ and $\mathcal L_{\mathrm{null}}$ are the summed Poisson negative log-likelihoods over the same evaluation entries for the model predictions and a homogeneous Poisson null using each neuron's training-set mean firing rate, respectively.
The denominator normalizes the likelihood improvement by the total observed spike count and converts it to bits.
Zero indicates the same Poisson score as this mean-firing-rate baseline.
We report the mean session score with SEM across sessions.
We additionally report the coefficient of determination ($R^2$) between predicted and observed binned counts as a supplementary metric.

\subsection{Baseline Models}
\label{sec:baselines}

We compare \ours{} with models that predict the future window in parallel or generate future activity sequentially from the observed neural history.
\textbf{IBL-MTM}~\citep{ibl2024mtm} masks the future bins and reconstructs them in parallel using the observed history as context.
\textbf{POCO}~\citep{duan2025poco} combines neuron-specific forecasting with population conditioning to predict the entire future window in a single forward pass.
\textbf{LFADS}~\citep{pandarinath2018inferring} infers an initial latent state from the observed history and evolves its generator autonomously over the future window, with controller and external inputs disabled.
\textbf{Causal NDT2}~\citep{ye2023ndt2} is adapted for next-bin count prediction and advances through the future window by feeding back counts sampled from its predictions.
\textbf{Neuroformer}~\citep{ye2023neuroformer} generates neural events autoregressively, which we convert to binned counts for evaluation.

In the current implementations, IBL-MTM and POCO are evaluated in Direct mode, and LFADS in RO mode.
\ours{}, causal NDT2, and Neuroformer are evaluated in both TF and RO modes, conditioning on ground-truth prefixes in TF and model-generated outputs in RO.
Model adaptations, feedback mechanisms, and scoring details are described in Appendix~\ref{app:baseline_adaptation}.

\subsection{Training Configuration}
\label{sec:implementation}

The default \ours{} model uses a 6-layer encoder with hidden size 128 and a 2-layer causal decoder.
Its prediction memory combines bin-level representations of the 4 most recent bins with summaries of the prediction prefix in segments of up to 5 bins.
We train \ours{} for 300 epochs using SparseLamb, bfloat16 mixed precision, and a Poisson negative log-likelihood objective.
Scheduled sampling linearly increases the probability of feeding model predictions into memory from 0 to 0.5 over the first 150 epochs.
Table~\ref{tab:hyperparams} lists the training hyperparameters; Appendix~\ref{app:training} provides training and compute details.

\begin{table}[!t]
\centering
\caption{\textbf{Forecasting performance across horizons.} Forward-prediction bits per spike (\fpbps{}) on the Perich--Miller dataset using 500\,ms of observed neural history. Direct and rollout (RO) forecasts use only this initial history, while teacher-forced (TF) evaluation receives true preceding activity. Values are session means with SEM. The highest mean in each column within each evaluation block is bold.}
\label{tab:main_results}
\footnotesize
\setlength{\tabcolsep}{4pt}
\begin{tabular}{lcccc}
\toprule
Model (mode) & 100\,ms & 240\,ms & 500\,ms & 1000\,ms \\
\midrule
\multicolumn{5}{l}{\textit{Autonomous forecasting}} \\
IBL-MTM (Direct) & 0.226\,{\tiny(0.019)} & 0.200\,{\tiny(0.013)} & 0.120\,{\tiny(0.011)} & -0.018\,{\tiny(0.025)} \\
POCO (Direct) & 0.177\,{\tiny(0.020)} & 0.136\,{\tiny(0.019)} & 0.070\,{\tiny(0.017)} & 0.035\,{\tiny(0.007)} \\
Neuroformer (RO) & -0.707\,{\tiny(0.059)} & -0.779\,{\tiny(0.066)} & -2.478\,{\tiny(0.241)} & -5.473\,{\tiny(0.117)} \\
LFADS (RO) & 0.212\,{\tiny(0.017)} & 0.204\,{\tiny(0.014)} & 0.172\,{\tiny(0.014)} & 0.139\,{\tiny(0.011)} \\
Causal NDT2 (RO) & 0.235\,{\tiny(0.019)} & 0.156\,{\tiny(0.015)} & 0.095\,{\tiny(0.017)} & 0.035\,{\tiny(0.021)} \\
\ours{} (RO) & \textbf{0.254}\,{\tiny(0.020)} & \textbf{0.224}\,{\tiny(0.016)} & \textbf{0.188}\,{\tiny(0.015)} & \textbf{0.143}\,{\tiny(0.011)} \\
\midrule
\multicolumn{5}{l}{\textit{TF evaluation}} \\
Neuroformer (TF) & -0.735\,{\tiny(0.053)} & -0.641\,{\tiny(0.049)} & -0.498\,{\tiny(0.068)} & -1.676\,{\tiny(0.389)} \\
Causal NDT2 (TF) & \textbf{0.296}\,{\tiny(0.022)} & \textbf{0.299}\,{\tiny(0.024)} & \textbf{0.289}\,{\tiny(0.022)} & 0.291\,{\tiny(0.025)} \\
\ours{} (TF) & 0.281\,{\tiny(0.024)} & 0.285\,{\tiny(0.024)} & 0.281\,{\tiny(0.025)} & \textbf{0.295}\,{\tiny(0.027)} \\
\bottomrule
\end{tabular}
\end{table}

\section{Results}
\label{sec:results}

\subsection{Autonomous Forecasting Across Horizons}
\label{sec:main_results}

Longer prediction horizons test how well models sustain forecasts using only an initial neural history.
We compare models on the Perich--Miller dataset using 500\,ms of observed activity and prediction horizons of 100, 240, 500, and 1000\,ms (Table~\ref{tab:main_results}).

Autonomous forecasting scores generally decrease as the horizon grows.
\ours{} achieves the highest mean at all four horizons, with rollout \fpbps{} of 0.254, 0.224, 0.188, and 0.143, respectively.
Session-level comparisons show that \ours{} remains above the mean-firing-rate baseline across all recordings and horizons, while differences between models vary across sessions (Table~\ref{tab:per_session_full}).
\ours{} also achieves the highest mean autonomous $R^2$ at each horizon, supporting its performance beyond the likelihood-based metric (Table~\ref{tab:r2}).
By contrast, TF scores for \ours{} and causal NDT2 remain relatively stable across horizons despite their declining rollout scores.
On the MC\_Maze forecasting adaptation, \ours{} also achieves higher autonomous \fpbps{} than IBL-MTM and POCO at 500\,ms (Table~\ref{tab:nlb_mc_maze}).
The example forecasts broadly reflect the observed differences in activity levels across neurons (Figure~\ref{fig:S1}).
A complementary Lorenz experiment examines information about known latent trajectories in the predicted activity.
\ours{} achieves slightly higher aligned latent $R^2$ and trajectory correlation than causal NDT2 (Table~\ref{tab:synthetic_lorenz} and Figure~\ref{fig:synthetic_lorenz}).

\subsection{Per-Bin Temporal Decay}
\label{sec:per_bin}

\begin{figure}[!ht]
    \centering
    \includegraphics[width=\linewidth]{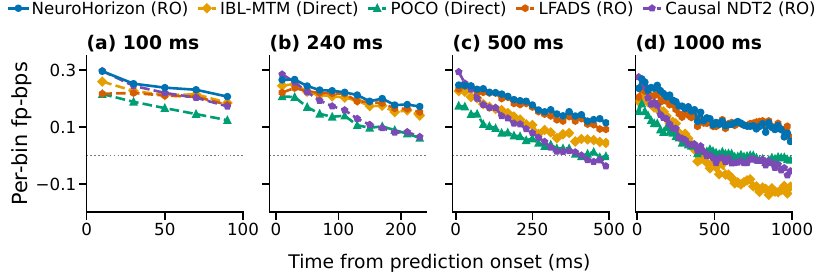}
    \caption{\textbf{Per-bin temporal decay of forecasting performance.} Panels (a--d) show per-bin \fpbps{} for prediction horizons of 100, 240, 500, and 1000\,ms, respectively, using a 500\,ms observation window. Scores use global spike weighting across sessions, neurons, and evaluation windows. All models forecast without additional neural observations, with prediction modes indicated in the legend.}
    \label{fig:per_bin}
\end{figure}

Whole-window scores do not reveal how prediction quality changes within a forecast.
We therefore examine per-bin \fpbps{} to determine how long predictions remain informative (Figure~\ref{fig:per_bin}).
Scores generally decline as prediction proceeds within each horizon.
In the later portion of the 1000\,ms window, IBL-MTM, POCO, and causal NDT2 reach near-zero or negative scores, indicating performance close to or below the mean-firing-rate baseline.

LFADS and \ours{} retain positive scores throughout that window, demonstrating predictive value beyond this baseline.
\ours{} has higher per-bin scores than LFADS throughout the 100, 240, and 500\,ms windows, complementing its higher whole-window means in Table~\ref{tab:main_results}.
Nevertheless, the positive scores of both models late in the 1000\,ms window are much lower than their early scores, indicating that the predictive gains captured by these models remain limited at longer delays.

\subsection{Ablation Study}
\label{sec:ablation}

\begin{table}[!t]
    \centering
    \caption{\textbf{Effects of prediction memory and scheduled sampling.} Values are session means of \fpbps{} with SEM. (a)~Four configurations compare no prediction memory, tail memory alone, segment memory alone, and their combination. Configurations with prediction memory use the same scheduled-sampling schedule with $\epsilon_{\max}\!=\!0.50$. TF and RO yield identical predictions without prediction memory, since neither true nor predicted counts are fed back. (b)~The combined tail and segment memory is retained while $\epsilon_{\max}$ is varied. The highest mean in each column within each section is bold.}
    \label{tab:ablation}
    \small
    \setlength{\tabcolsep}{5.5pt}
    \begin{tabular}{l cccc}
        \toprule
        & \multicolumn{2}{c}{\textbf{500\,ms}} & \multicolumn{2}{c}{\textbf{1000\,ms}} \\
        \cmidrule(lr){2-3} \cmidrule(lr){4-5}
        \textbf{Configuration} & TF & RO & TF & RO \\
        \midrule
        \multicolumn{5}{l}{\textit{(a) Prediction memory}} \\
        \quad No memory & 0.174\,{\tiny(0.014)} & 0.174\,{\tiny(0.014)} & 0.139\,{\tiny(0.012)} & 0.139\,{\tiny(0.012)} \\
        \quad Only tail memory & 0.278\,{\tiny(0.024)} & 0.182\,{\tiny(0.015)} & 0.291\,{\tiny(0.026)} & 0.126\,{\tiny(0.013)} \\
        \quad Only segment memory & 0.268\,{\tiny(0.025)} & 0.169\,{\tiny(0.015)} & 0.274\,{\tiny(0.027)} & 0.126\,{\tiny(0.012)} \\
        \quad \textbf{\ours{}} (tail and segment memory) & \textbf{0.281}\,{\tiny(0.025)} & \textbf{0.188}\,{\tiny(0.015)} & \textbf{0.295}\,{\tiny(0.027)} & \textbf{0.143}\,{\tiny(0.011)} \\
        \midrule
        \multicolumn{5}{l}{\textit{(b) Scheduled sampling}} \\
        \quad No SS ($\epsilon\!=\!0$) & \textbf{0.293}\,{\tiny(0.024)} & 0.073\,{\tiny(0.025)} & \textbf{0.296}\,{\tiny(0.023)} & $-$0.123\,{\tiny(0.022)} \\
        \quad Low SS ($\epsilon_{\max}\!=\!0.25$) & 0.291\,{\tiny(0.023)} & 0.180\,{\tiny(0.013)} & \textbf{0.296}\,{\tiny(0.025)} & 0.128\,{\tiny(0.012)} \\
        \quad \textbf{\ours{}} ($\epsilon_{\max}\!=\!0.50$) & 0.281\,{\tiny(0.025)} & \textbf{0.188}\,{\tiny(0.015)} & 0.295\,{\tiny(0.027)} & \textbf{0.143}\,{\tiny(0.011)} \\
        \bottomrule
    \end{tabular}
\end{table}

Long-horizon rollout depends on how the decoder represents its prediction history and how training exposes it to predicted inputs.
We assess these two aspects by comparing prediction-memory configurations and scheduled-sampling settings at 500 and 1000\,ms under TF and RO evaluation (Table~\ref{tab:ablation}).

\textbf{Prediction memory.}
To examine the role of temporal resolution, we compare no memory, tail memory alone, segment memory alone, and their combination (Table~\ref{tab:ablation}(a)).
The combined memory achieves the highest rollout means at both horizons, exceeding either single-memory configuration and the no-memory scores of 0.174 and 0.139.
Its advantage over tail memory alone is larger at 1000\,ms.
These results support combining recent bin-level detail with segment summaries of the prediction prefix.

Separate sensitivity analyses examine observed history and generated-history tail capacity at a 1000\,ms prediction horizon.
Increasing observation history from 0.5 to 1 and 2\,s does not yield monotonic gains, so additional observations do not automatically improve rollout under these settings (Table~\ref{tab:observation_history_sensitivity}).
Reducing $W_{\mathrm{tail}}$ from 4 to 1 lowers performance, while increasing it to 8 does not improve it, indicating that a longer tail is not necessarily more effective (Table~\ref{tab:tail_capacity_sensitivity}).

\textbf{Scheduled sampling.}
Training on true prefixes differs from conditioning on predictions during rollout.
We examine the effect of this mismatch by varying $\epsilon_{\max}\in\{0,0.25,0.50\}$ with the combined memory fixed (Table~\ref{tab:ablation}(b)).
Without scheduled sampling, rollout \fpbps{} is 0.073 at 500\,ms and $-0.123$ at 1000\,ms.
Scheduled sampling improves rollout at both horizons, with $\epsilon_{\max}\!=\!0.50$ yielding the highest means of 0.188 and 0.143.
TF scores remain comparatively similar across settings.
The greater improvement in rollout supports exposing the decoder to its own predictions during training to sustain autonomous forecasting.

\subsection{Scaling and Cross-Subject Transfer}
\label{sec:scaling_generalization}

\begin{figure}[!t]
    \centering
    \includegraphics[width=\linewidth]{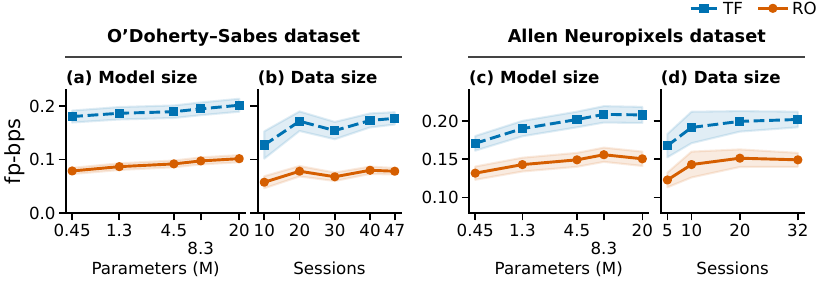}
    \caption{\textbf{Scaling analysis.} Panels (a,b) show results on the O'Doherty--Sabes dataset, and panels (c,d) show results on the Allen Neuropixels dataset. Model-size experiments use 47 and 32 recording sessions, respectively. Data-size experiments vary the number of training sessions and evaluate the sessions included at each size. All panels report session means with $\pm 1$ SEM for TF and RO \fpbps{}, using 500\,ms observation and prediction windows. Parameter counts are shown in millions and exclude neuron and session embeddings.}
    \label{fig:scaling}
\end{figure}

\begin{table}[!t]
    \centering
    \caption{\textbf{Cross-subject transfer with full fine-tuning.} Results on the O'Doherty--Sabes dataset of macaque sensorimotor recordings, with Indy as the source subject and Loco as the target subject. Pretraining on 37 Indy sessions followed by full fine-tuning on the training splits of five Loco sessions is compared with training from scratch on the same target data. Values are test-session means of \fpbps{} with SEM across the five Loco sessions. The highest mean in each column is bold.}
    \label{tab:generalization}
    \small
    \setlength{\tabcolsep}{6pt}
    \begin{tabular}{l cc cc}
        \toprule
        & \multicolumn{2}{c}{\textbf{500\,ms}} & \multicolumn{2}{c}{\textbf{1000\,ms}} \\
        \cmidrule(lr){2-3} \cmidrule(lr){4-5}
        \textbf{Setting} & TF & RO & TF & RO \\
        \midrule
        From scratch (5 Loco) & 0.058\,{\tiny(0.008)} & 0.026\,{\tiny(0.007)} & 0.060\,{\tiny(0.010)} & \textbf{0.018}\,{\tiny(0.008)} \\
        Transfer full fine-tune (37 Indy $\to$ 5 Loco) & \textbf{0.082}\,{\tiny(0.008)} & \textbf{0.035}\,{\tiny(0.009)} & \textbf{0.076}\,{\tiny(0.010)} & 0.015\,{\tiny(0.010)} \\
        \bottomrule
    \end{tabular}
\end{table}

Extending forecasting to broader recording settings raises questions about model capacity, training data size, and initialization for new subjects.
We examine scaling on two datasets and transfer through cross-subject pretraining and fine-tuning.

\textbf{Capacity and data scaling.}
To assess how forecasting changes with capacity and available recordings, we vary model size and the number of training sessions on the O'Doherty--Sabes dataset and the Allen Neuropixels dataset, using 500\,ms observation and prediction windows (Figure~\ref{fig:scaling}).
The five model configurations span 0.45--20.00 million backbone parameters (Table~\ref{tab:param_scaling}).
Both datasets show overall improvements in TF and RO performance as model and data size increase, with smaller gains at larger scales on the Allen Neuropixels dataset.
These trends motivate further study of model capacity and recording scale as ways to improve forecasting.

\textbf{Cross-subject transfer.}
Pretraining on another subject may provide a useful initialization for a new neural population.
We test this by comparing pretraining on 37 Indy sessions followed by full fine-tuning on five Loco sessions with training from scratch on the same target data (Table~\ref{tab:generalization}).
The architecture and target training settings are shared.
Pretraining improves the TF and RO point estimates at 500\,ms, and the TF improvement persists at 1000\,ms.
The 500\,ms results indicate that source-subject pretraining can benefit autonomous prediction after adaptation to the target subject.
\FloatBarrier

\section{Discussion}
\label{sec:discussion}

We investigated how long neural population activity remains predictable from an initial spike history without further observations, comparing direct multi-step prediction and autoregressive rollout across horizons through whole-window scores and per-bin prediction decay.
\ours{}, which combines hierarchical prediction memory with scheduled sampling, achieves the highest mean autonomous forecasting scores across the evaluated horizons on the Perich--Miller dataset and retains positive per-bin scores throughout the one-second window (Table~\ref{tab:main_results}; Figure~\ref{fig:per_bin}).
Its advantage narrows at 1000\,ms, and late-window scores remain well below early-window scores.
Ablations show that rollout benefits from combining tail and segment memory and degrades sharply without scheduled sampling (Table~\ref{tab:ablation}).
Together with the scaling and transfer results, these findings indicate that motor-cortical population activity remains partially predictable for up to one second from initial spike history, and that prediction memory and scheduled sampling improve autonomous rollout.

Future work could examine how far autonomous forecasting extends and what information supports it at longer horizons.
Evaluating larger models and datasets over longer horizons, and over a broader range of observation durations, would test whether the observed scaling trends hold beyond the evaluated windows.
Analyzing learned representations, incorporating stimulus inputs, and relating predicted activity to behavior could further reveal which aspects of population dynamics remain predictive and whether forecasts benefit from information beyond spike history.

\begin{ack}
This work was supported by a grant from the National Natural Science Foundation of China (No.~92370116), the Beijing Major Science and Technology Project (Contract No.~Z251100008125058), and the Tsinghua University Dushi Program. The authors declare no competing interests.
\end{ack}

\bibliographystyle{unsrtnat}
\bibliography{references}

\newpage
\appendix
\renewcommand{\thefigure}{S\arabic{figure}}
\setcounter{figure}{0}
\renewcommand{\thetable}{S\arabic{table}}
\setcounter{table}{0}
\setcounter{topnumber}{3}
\setcounter{totalnumber}{4}
\renewcommand{\topfraction}{0.9}

\section{Dataset Splits}
\label{app:data_splits}

We retain the training, validation, and test domains generated by the brainsets preprocessing pipelines for the Perich--Miller and O'Doherty--Sabes datasets.
Perich--Miller uses trial-based partitions with buffers around held-out intervals, while O'Doherty--Sabes assigns ten temporal blocks to eight training blocks and one block each for validation and test, following the pipeline's movement-based allocation and filtering rules.
For the Allen Neuropixels dataset, complete repeats of natural movie one and natural movie three are assigned chronologically to the three splits.
We allocate 14 repeats of natural movie one to training, two to validation, and four to testing.
For natural movie three, the corresponding numbers are seven, two, and one.
\FloatBarrier

\section{Baseline Model Adaptations}
\label{app:baseline_adaptation}

We adapt the baseline models to forecasting from an initial neural history and evaluate their binned count predictions with the shared Poisson score and training-set null model.

\textbf{IBL-MTM.}
We retain the self-supervised masking objective of IBL-MTM~\citep{ibl2024mtm}, alternating between neuron masks and causal masks with a base mask ratio of $0.3$.
Neuron masks hide subsets of neurons across the input window, while causal masks hide contiguous time spans.
For Direct evaluation, the observed history provides context and all future target bins are masked and reconstructed jointly.

\textbf{POCO.}
We retain the upstream POCO model core~\citep{duan2025poco} and adapt its inputs and outputs to binned spike-count forecasting with a Poisson negative log-likelihood objective.
The model receives only the observed history and predicts the entire future window in one forward pass, so we report it in Direct mode.

\textbf{LFADS.}
We use the \texttt{lfads-torch} implementation of LFADS~\citep{pandarinath2018inferring} in a multisession model with session-specific input and output mappings.
Its encoder receives only the 500\,ms observation counts, and the generator evolves autonomously over the future window with controller and external inputs disabled.
The initial-condition latent and factors are 64-dimensional, and the generator state is 128-dimensional.
Training combines a Poisson reconstruction objective over the history and future with initial-condition KL and recurrent-weight regularization, with checkpoints selected by validation ELBO.
Future counts serve only as training targets, and forecasting is evaluated in RO mode.

\textbf{Causal NDT2.}
We adapt the official multisession NDT2 model~\citep{ye2023ndt2} from masked reconstruction to next-bin prediction under causal attention.
Counts from bin $t-1$ are encoded at query time $t$, where a learned mask token queries bin $t$ and the true target counts enter only the Poisson loss.
The model is trained from scratch for 300 epochs with an effective batch size of 32 and a 500\,ms observation window.
Training and TF evaluation condition on true preceding counts, whereas RO feeds back counts sampled from the predicted Poisson distributions without access to true future activity.

\textbf{Neuroformer.}
We retain the causal event-transformer components of Neuroformer~\citep{ye2023neuroformer}.
Its feature, history, and state attention modules each contain six blocks.
The model uses eight attention heads with hidden size 256 and dropout 0.2.
History event blocks contain up to 512 events and current blocks contain up to 256, with 10\,ms time tokens.
Because the model generates events rather than a matrix of expected bin counts, we sample event sequences and assign their events to bins using the generated timestamps.
For each evaluation window, we average the binned counts from 20 sampled sequences and apply a $10^{-4}$ floor before computing the Poisson count score.
In TF evaluation, each event prediction conditions on the true event prefix, whereas RO incorporates generated events into the context for subsequent predictions.
\FloatBarrier

\section{Training Configuration and Compute}
\label{app:training}

The training settings for \ours{} on the Perich--Miller dataset are summarized in Table~\ref{tab:hyperparams}.
Training uses bfloat16 mixed precision on a single NVIDIA RTX~4090 or RTX~4080 SUPER GPU.
The 300-epoch runs take roughly 2--3 hours at 100\,ms, 3.5 hours at 240\,ms, 8--9 hours at 500\,ms, and 17--18 hours at 1000\,ms.
On the RTX~4090 with 24\,GB of memory, training at batch size 32 peaks at approximately 12--16\,GB.
TF and RO evaluation over the corresponding test set each complete in less than one hour.

\begin{table}[!ht]
    \centering
    \caption{\textbf{Training hyperparameters.} Values correspond to the NeuroHorizon configuration used for experiments on the Perich--Miller dataset.}
    \label{tab:hyperparams}
    \footnotesize
    \setlength{\tabcolsep}{3pt}
    \renewcommand{\arraystretch}{1.1}
    \begin{tabular}{@{}>{\raggedright\arraybackslash}p{0.29\linewidth}>{\raggedright\arraybackslash}p{0.15\linewidth}@{\hspace{1.5em}}>{\raggedright\arraybackslash}p{0.19\linewidth}>{\raggedright\arraybackslash}p{0.295\linewidth}@{}}
        \toprule
        \textbf{Hyperparameter} & \textbf{Value} & \textbf{Hyperparameter} & \textbf{Value} \\
        \midrule
        \multicolumn{2}{@{}l}{\textit{Encoder (POYO+)}} & \multicolumn{2}{@{}l@{}}{\textit{Training}} \\
        Number of layers ($L_{\text{enc}}$) & 6 & Optimizer & SparseLamb \\
        Hidden dimension ($d$) & 128 & Peak learning rate & $3.125 \times 10^{-5} \times$ batch size \\
        Attention heads & 8 self, 2 cross & LR schedule & OneCycleLR with cosine annealing \\
        Head dimension ($d_{\text{head}}$) & 64 & Weight decay & $10^{-4}$ \\
        Dropout (attention, FFN) & 0.2 & Batch size & 32 \\
        Dropout (input) & 0.4 & Epochs & 300 \\
        Latent step & 50\,ms & & \\
        Latents per step & 32 & & \\
        \multicolumn{2}{@{}l}{\textit{Decoder}} & \multicolumn{2}{@{}l@{}}{\textit{Scheduled Sampling}} \\
        Number of layers ($L_{\text{dec}}$) & 2 & Schedule & Linear ramp, $\epsilon: 0 \to 0.5$ \\
        Hidden dimension ($d$) & 128 & Ramp duration & 150 epochs \\
        Tokens per tail bin ($K_{\mathrm{tail}}$) & 4 & \multicolumn{2}{@{}l@{}}{\textit{Memory Robustness}} \\
        Prediction memory attention heads & 4 & Input noise ($\sigma_{\text{noise}}$) & 0.03 \\
        Tail memory window ($W_{\mathrm{tail}}$) & 4 bins (80\,ms) & Input dropout ($p_{\text{in}}$) & 0.05 \\
        Segment chunk size ($S$) & 5 bins (100\,ms) & Block dropout ($p_{\text{block}}$) & 0.15 \\
        Segment tokens per chunk ($K_{\text{seg}}$) & 1 & & \\
        \bottomrule
    \end{tabular}
\end{table}

\section{Session-Level Forecasting Results}
\label{app:per_session}

To examine how the overall forecasting scores reflect individual recordings, we compare session-level performance across the four prediction horizons on the Perich--Miller dataset in Table~\ref{tab:per_session_full}.
\ours{} maintains positive rollout \fpbps{} in every session and at every horizon, and exceeds both IBL-MTM and POCO in all 40 session--horizon comparisons.
These results indicate that the higher mean scores relative to the direct baselines reflect improvements across recordings and prediction horizons.

\begin{table}[!htbp]
    \centering
    \caption{\textbf{Per-session forward-prediction results (\fpbps{}).} C, J, and M denote Chewie, Jango, and Mihili, the three monkeys in the Perich--Miller dataset. Sessions are numbered chronologically within each monkey. Each horizon separates autonomous forecasts (Direct and RO) from TF evaluations. The final column reports the session mean (SEM).}
    \label{tab:per_session_full}
    \footnotesize
    \setlength{\tabcolsep}{1pt}
    \begin{tabular}{@{}ll*{11}{r}@{}}
        \toprule
        \shortstack{\textbf{Horizon}\\\textbf{(ms)}} & \textbf{Model (mode)} & \textbf{C1} & \textbf{C2} & \textbf{C3} & \textbf{C4} & \textbf{J1} & \textbf{J2} & \textbf{J3} & \textbf{M1} & \textbf{M2} & \textbf{M3} & \textbf{Mean (SEM)} \\
        \midrule
        \multirow{7}{*}{100} & IBL-MTM (Direct) & 0.171 & 0.147 & 0.139 & 0.198 & 0.266 & 0.290 & 0.299 & 0.273 & 0.243 & 0.237 & 0.226 (0.019) \\
         & POCO (Direct) & 0.133 & 0.104 & 0.091 & 0.147 & 0.245 & 0.254 & 0.268 & 0.163 & 0.168 & 0.199 & 0.177 (0.020) \\
         & LFADS (RO) & 0.163 & 0.159 & 0.121 & 0.186 & 0.264 & 0.275 & 0.262 & 0.253 & 0.207 & 0.225 & 0.212 (0.017) \\
         & Causal NDT2 (RO) & 0.176 & 0.165 & 0.144 & 0.193 & 0.270 & 0.297 & 0.303 & 0.269 & 0.270 & 0.259 & 0.235 (0.019) \\
         & \ours{} (RO) & 0.192 & 0.185 & 0.156 & 0.222 & 0.294 & 0.345 & 0.331 & 0.283 & 0.267 & 0.262 & 0.254 (0.020) \\
        \cmidrule(lr){2-13}
         & Causal NDT2 (TF) & 0.222 & 0.219 & 0.204 & 0.237 & 0.364 & 0.385 & 0.376 & 0.322 & 0.324 & 0.310 & 0.296 (0.022) \\
         & \ours{} (TF) & 0.212 & 0.199 & 0.170 & 0.234 & 0.358 & 0.379 & 0.378 & 0.299 & 0.291 & 0.288 & 0.281 (0.024) \\
        \midrule
        \multirow{7}{*}{240} & IBL-MTM (Direct) & 0.185 & 0.153 & 0.121 & 0.184 & 0.224 & 0.251 & 0.248 & 0.229 & 0.207 & 0.194 & 0.200 (0.013) \\
         & POCO (Direct) & 0.140 & 0.110 & $-$0.016 & 0.143 & 0.170 & 0.170 & 0.216 & 0.127 & 0.159 & 0.142 & 0.136 (0.019) \\
         & LFADS (RO) & 0.180 & 0.148 & 0.132 & 0.201 & 0.207 & 0.269 & 0.256 & 0.238 & 0.211 & 0.200 & 0.204 (0.014) \\
         & Causal NDT2 (RO) & 0.127 & 0.123 & 0.067 & 0.134 & 0.192 & 0.138 & 0.197 & 0.222 & 0.186 & 0.178 & 0.156 (0.015) \\
         & \ours{} (RO) & 0.189 & 0.173 & 0.144 & 0.201 & 0.241 & 0.299 & 0.299 & 0.240 & 0.249 & 0.208 & 0.224 (0.016) \\
        \cmidrule(lr){2-13}
         & Causal NDT2 (TF) & 0.228 & 0.235 & 0.198 & 0.242 & 0.373 & 0.427 & 0.370 & 0.320 & 0.302 & 0.297 & 0.299 (0.024) \\
         & \ours{} (TF) & 0.226 & 0.216 & 0.176 & 0.236 & 0.369 & 0.399 & 0.383 & 0.288 & 0.292 & 0.267 & 0.285 (0.024) \\
        \midrule
        \multirow{7}{*}{500} & IBL-MTM (Direct) & 0.128 & 0.096 & 0.075 & 0.132 & 0.125 & 0.142 & 0.183 & 0.155 & 0.092 & 0.077 & 0.120 (0.011) \\
         & POCO (Direct) & 0.079 & 0.066 & 0.028 & 0.082 & 0.063 & 0.169 & 0.134 & 0.039 & $-$0.028 & 0.066 & 0.070 (0.017) \\
         & LFADS (RO) & 0.145 & 0.121 & 0.102 & 0.152 & 0.202 & 0.204 & 0.235 & 0.234 & 0.171 & 0.157 & 0.172 (0.014) \\
         & Causal NDT2 (RO) & 0.080 & 0.043 & 0.002 & 0.057 & 0.092 & 0.097 & 0.160 & 0.172 & 0.137 & 0.109 & 0.095 (0.017) \\
         & \ours{} (RO) & 0.151 & 0.140 & 0.115 & 0.157 & 0.205 & 0.247 & 0.258 & 0.234 & 0.191 & 0.177 & 0.188 (0.015) \\
        \cmidrule(lr){2-13}
         & Causal NDT2 (TF) & 0.220 & 0.210 & 0.191 & 0.231 & 0.368 & 0.363 & 0.364 & 0.329 & 0.310 & 0.304 & 0.289 (0.022) \\
         & \ours{} (TF) & 0.211 & 0.202 & 0.182 & 0.218 & 0.367 & 0.392 & 0.380 & 0.298 & 0.277 & 0.285 & 0.281 (0.025) \\
        \midrule
        \multirow{7}{*}{1000} & IBL-MTM (Direct) & 0.116 & 0.028 & 0.025 & 0.022 & $-$0.079 & $-$0.072 & $-$0.029 & 0.043 & $-$0.081 & $-$0.154 & $-$0.018 (0.025) \\
         & POCO (Direct) & 0.063 & 0.037 & 0.002 & 0.050 & 0.044 & 0.049 & 0.066 & 0.017 & 0.017 & 0.004 & 0.035 (0.007) \\
         & LFADS (RO) & 0.137 & 0.112 & 0.087 & 0.124 & 0.121 & 0.151 & 0.208 & 0.174 & 0.142 & 0.138 & 0.139 (0.011) \\
         & Causal NDT2 (RO) & 0.049 & 0.008 & $-$0.012 & 0.058 & 0.075 & $-$0.113 & 0.121 & 0.015 & 0.087 & 0.063 & 0.035 (0.021) \\
         & \ours{} (RO) & 0.165 & 0.129 & 0.105 & 0.140 & 0.118 & 0.130 & 0.191 & 0.207 & 0.128 & 0.114 & 0.143 (0.011) \\
        \cmidrule(lr){2-13}
         & Causal NDT2 (TF) & 0.219 & 0.220 & 0.177 & 0.229 & 0.354 & 0.389 & 0.406 & 0.294 & 0.317 & 0.307 & 0.291 (0.025) \\
         & \ours{} (TF) & 0.225 & 0.216 & 0.183 & 0.231 & 0.356 & 0.420 & 0.421 & 0.312 & 0.296 & 0.292 & 0.295 (0.027) \\
        \bottomrule
    \end{tabular}
\end{table}

\section[Evaluation with R2]{Evaluation with $R^2$}
\label{app:supplementary_metrics}

To complement the likelihood-based score, we evaluate how closely predicted expected counts match observed binned counts on the same Perich--Miller test recordings.
For each session $s$, we pool all valid evaluation windows, prediction bins, and neurons to compute
\begin{equation}
    R_s^2=1-\frac{\sum_{w,\tau,n}(\hat\mu_{w,\tau,n}-y_{w,\tau,n})^2}
    {\sum_{w,\tau,n}(y_{w,\tau,n}-\bar y_s)^2},
    \label{eq:session_r2}
\end{equation}
where $\bar y_s$ is the mean observed count over those entries within session $s$.
We then report the mean and SEM across sessions, rather than averaging separate neuron-level $R^2$ values.
\ours{} achieves the highest mean autonomous $R^2$ at each horizon in Table~\ref{tab:r2}, complementing the comparison based on \fpbps{}.
TF results use additional true-prefix information and are reported separately.

\begin{table}[!ht]
\centering
\caption{\textbf{Spike-count prediction accuracy.} Session means of $R^2$ with SEM on the Perich--Miller dataset. Autonomous forecasts use only the initial history, while TF evaluation receives true prefixes.}
\label{tab:r2}
\small
\begin{tabular}{lcccc}
\toprule
Model (mode) & 100\,ms & 240\,ms & 500\,ms & 1000\,ms \\
\midrule
\multicolumn{5}{l}{\textit{Autonomous forecasting}} \\
IBL-MTM (Direct) & 0.245\,{\tiny(0.017)} & 0.237\,{\tiny(0.018)} & 0.211\,{\tiny(0.019)} & 0.175\,{\tiny(0.018)} \\
POCO (Direct) & 0.230\,{\tiny(0.017)} & 0.217\,{\tiny(0.018)} & 0.200\,{\tiny(0.019)} & 0.183\,{\tiny(0.017)} \\
Neuroformer (RO) & 0.128\,{\tiny(0.018)} & 0.096\,{\tiny(0.014)} & -2.722\,{\tiny(1.341)} & -1.079\,{\tiny(0.449)} \\
LFADS (RO) & 0.242\,{\tiny(0.017)} & 0.238\,{\tiny(0.018)} & 0.226\,{\tiny(0.019)} & 0.209\,{\tiny(0.016)} \\
Causal NDT2 (RO) & 0.248\,{\tiny(0.017)} & 0.217\,{\tiny(0.018)} & 0.197\,{\tiny(0.015)} & 0.174\,{\tiny(0.021)} \\
\ours{} (RO) & 0.255\,{\tiny(0.018)} & 0.244\,{\tiny(0.018)} & 0.232\,{\tiny(0.018)} & 0.210\,{\tiny(0.017)} \\
\midrule
\multicolumn{5}{l}{\textit{TF evaluation}} \\
Neuroformer (TF) & 0.142\,{\tiny(0.014)} & 0.145\,{\tiny(0.016)} & 0.173\,{\tiny(0.022)} & 0.116\,{\tiny(0.015)} \\
Causal NDT2 (TF) & 0.274\,{\tiny(0.018)} & 0.274\,{\tiny(0.020)} & 0.270\,{\tiny(0.021)} & 0.270\,{\tiny(0.018)} \\
\ours{} (TF) & 0.266\,{\tiny(0.019)} & 0.268\,{\tiny(0.021)} & 0.267\,{\tiny(0.021)} & 0.271\,{\tiny(0.018)} \\
\bottomrule
\end{tabular}
\end{table}

\section{Forecasting on the MC\_Maze Dataset}
\label{app:nlb_mc_maze}

To assess forecasting beyond the Perich--Miller recordings, we compare \ours{}, IBL-MTM, and POCO on the single MC\_Maze recording from the Neural Latents Benchmark~\citep{pei2021neural}.
Each model observes 500\,ms of neural activity and forecasts the next 500\,ms using the same data splits and evaluation windows.
The official training trials are divided into local training and validation sets, while the official validation trials form the local test set.
This is a forecasting adaptation of MC\_Maze, not an official NLB benchmark score.
\ours{} achieves higher autonomous \fpbps{} than both direct baselines, providing complementary evidence of forecasting performance under a different recording condition.

\begin{table}[!ht]
    \centering
    \caption{\textbf{Forecasting on MC\_Maze.} Global \fpbps{} and $R^2$ for 500\,ms forecasts from 500\,ms of observed activity. Autonomous forecasts and TF evaluation are reported separately.}
    \label{tab:nlb_mc_maze}
    \small
    \begin{tabular}{lcc}
        \toprule
        \textbf{Model (mode)} & \textbf{Global \fpbps{}} & \textbf{$R^2$} \\
        \midrule
        \multicolumn{3}{l}{\textit{Autonomous forecasting}} \\
        IBL-MTM (Direct) & 0.2475 & 0.1067 \\
        POCO (Direct) & 0.2162 & 0.1036 \\
        \ours{} (RO) & 0.2722 & 0.1114 \\
        \midrule
        \multicolumn{3}{l}{\textit{TF evaluation}} \\
        \ours{} (TF) & 0.3357 & 0.1226 \\
        \bottomrule
    \end{tabular}
\end{table}

\section{Qualitative Population Forecasts}
\label{app:visualizations}

\begin{figure}[!ht]
    \centering
    \includegraphics[width=\linewidth]{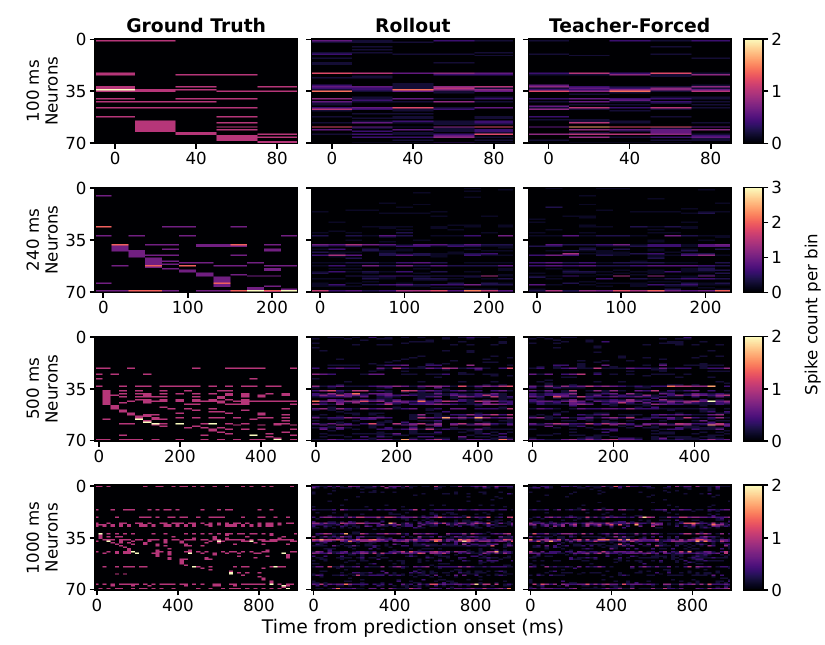}
    \caption{\textbf{Population forecasts across horizons.} Rows show 100, 240, 500, and 1000\,ms forecasts for 71 neurons. Columns show observed counts, rollout predictions, and teacher-forced predictions. Colors indicate counts per bin.}
    \label{fig:S1}
\end{figure}

To visualize population activity beyond the aggregate scores, Figure~\ref{fig:S1} compares observed counts with \ours{} predictions under RO and TF across the four horizons.
For display, each predicted expected count $\hat\mu$ is converted to the mean of five samples from $\operatorname{Poisson}(\hat\mu)$.
This sampling is used only for visualization, while autonomous rollout feeds back predicted expected counts.
The examples show broad correspondence in activity levels across neurons alongside differences in individual spike counts.
Quantitative comparisons rely on the forecasting scores and per-bin analysis.

\FloatBarrier

\section{Forecasting on Lorenz-Driven Spike Trains}
\label{app:synthetic_lorenz}

We use a controlled Lorenz system to test whether autonomous neural forecasts retain information about known latent trajectories.
We discard the initial 2\,s of each independent three-dimensional trajectory to reduce initial transients.
The latent coordinates are standardized using training-set statistics and mapped through a fixed nonlinear readout to the firing rates of a 64-unit Poisson population sampled in 20\,ms bins.
We compare \ours{} with causal NDT2 on this synthetic dataset.
Both models use the same 4,200 trajectories for training, 900 for validation, and 900 for test.
Each observes 500\,ms of spikes and autonomously forecasts the next 500\,ms.

\begin{figure}[!ht]
    \centering
    \includegraphics[width=4.0in]{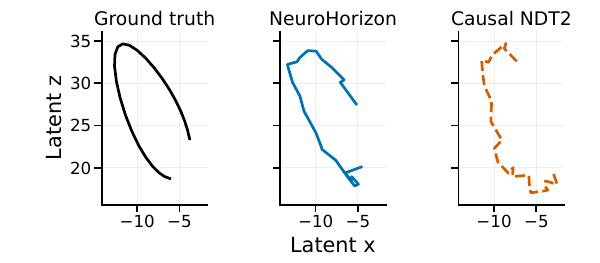}
    \caption{\textbf{Example Lorenz trajectories over a 500\,ms forecast.} A held-out trajectory and validation-aligned predictions from \ours{} and causal NDT2, projected onto the $x$--$z$ plane.}
    \label{fig:synthetic_lorenz}
\end{figure}

Table~\ref{tab:synthetic_lorenz} reports spike forecasting relative to a training-split Poisson null (fp-bps), recovery of the known generating rates (Rate $R^2$), and two latent-trajectory measures.
For each model, a ridge-regularized affine map from predicted log firing rates to ground-truth Lorenz coordinates is fitted on the validation set and frozen for test evaluation.
Latent $R^2$ compares these aligned predictions with the test latents. Trajectory correlation is the Pearson correlation after flattening the test trajectories and their coordinates.
Figure~\ref{fig:synthetic_lorenz} illustrates the aligned trajectories.
\ours{} has slightly higher latent $R^2$ and trajectory correlation, while causal NDT2 has higher fp-bps and Rate $R^2$.
These results indicate that the forecasts retain information about the underlying latent trajectories, with neither model leading on all four metrics.

\begin{table}[!ht]
    \centering
    \caption{\textbf{Forecasting on Lorenz-driven spike trains.} Autonomous 500\,ms forecasts after a 500\,ms observation. Metrics are evaluated on held-out trajectories, with latent alignment fitted on validation predictions.}
    \label{tab:synthetic_lorenz}
    \small
    \setlength{\tabcolsep}{6pt}
    \begin{tabular}{lcccc}
        \toprule
        \textbf{Model} & \textbf{fp-bps} & \textbf{Rate $R^2$} & \textbf{Latent $R^2$} & \textbf{Trajectory corr.} \\
        \midrule
        \ours{} & 0.1368 & 0.5535 & 0.7225 & 0.9517 \\
        Causal NDT2 & 0.1595 & 0.6835 & 0.7078 & 0.9491 \\
        \bottomrule
    \end{tabular}
\end{table}

\FloatBarrier

\section{Sensitivity to Observation History and Tail Length}
\label{app:history_tail_sensitivity}

\paragraph{Observation history.}
To examine whether additional observations improve long-horizon forecasting, we compare 0.5, 1, and 2\,s of observed neural history on the Perich--Miller dataset with a 1000\,ms prediction horizon (Table~\ref{tab:observation_history_sensitivity}).
The default setting uses 0.5\,s of history, and models for 1\,s and 2\,s are trained from scratch.
All three settings use 300 training epochs.
Extending observation history does not produce a monotonic rollout improvement under these settings.

\begin{table}[!ht]
    \centering
    \caption{\textbf{Observation-history sensitivity.} Autonomous 1000\,ms prediction on the Perich--Miller dataset. Rollout \fpbps{} is reported as mean $\pm$ SEM across sessions. $\Delta$ is the mean difference from the 0.5\,s reference.}
    \label{tab:observation_history_sensitivity}
    \small
    \begin{tabular}{lcc}
        \toprule
        \textbf{Observation history} & \textbf{Rollout \fpbps{}} & \textbf{$\Delta$ vs. 0.5\,s} \\
        \midrule
        0.5\,s & $0.1427 \pm 0.0107$ & --- \\
        1.0\,s & $0.1442 \pm 0.0172$ & $+0.0015$ \\
        2.0\,s & $0.1381 \pm 0.0317$ & $-0.0046$ \\
        \bottomrule
    \end{tabular}
\end{table}

\paragraph{Tail length.}
To assess how much recent prediction history to retain, we use 500\,ms of observed activity and vary the number $W_{\mathrm{tail}}$ of recent generated bins retained at high resolution while keeping the segment-memory pathway fixed (Table~\ref{tab:tail_capacity_sensitivity}).
Each tail bin contributes four memory tokens. Segment summaries use one token per five-bin group.
At the final prediction step, the decoder sees $4W_{\mathrm{tail}}+10$ prediction-history memory tokens, including the tail and segment summaries.
Reducing the tail from $W_{\mathrm{tail}}=4$ to $W_{\mathrm{tail}}=1$ lowers rollout \fpbps{} by 0.0148 on average across sessions. Increasing it to $W_{\mathrm{tail}}=8$ does not improve performance.
Among the tested configurations, $W_{\mathrm{tail}}=4$ gives the highest mean rollout and last-five-bin scores.

\begin{table}[!ht]
    \centering
    \caption{\textbf{Generated-history tail sensitivity.} Perich--Miller dataset, 500\,ms observation and 1000\,ms autonomous prediction, with the segment-memory configuration fixed. Visible tokens count prediction-history memory tokens available at the final query. Rollout \fpbps{} is the session mean $\pm$ SEM. Last-five-bin \fpbps{} is the arithmetic mean of the five global per-bin scores covering 900--1000\,ms (20\,ms bins).}
    \label{tab:tail_capacity_sensitivity}
    \small
    \begin{tabular}{cccc}
        \toprule
        \textbf{Tail $W_{\mathrm{tail}}$} & \textbf{Visible tokens} & \textbf{Rollout \fpbps{}} & \textbf{Last-five-bin \fpbps{}} \\
        \midrule
        1 & 14 & $0.1279 \pm 0.0117$ & 0.0561 \\
        4 & 26 & $0.1427 \pm 0.0107$ & 0.0770 \\
        8 & 42 & $0.1297 \pm 0.0112$ & 0.0720 \\
        \bottomrule
    \end{tabular}
\end{table}
\FloatBarrier

\section{Model Configurations for Scaling Experiments}
\label{app:scaling}

Table~\ref{tab:param_scaling} defines the five model sizes used in Figure~\ref{fig:scaling} for the O'Doherty--Sabes and Allen Neuropixels datasets.
The configurations vary hidden width, encoder and decoder depth, and attention configuration together.

\begin{table}[!ht]
    \centering
    \caption{\textbf{Parameter-scaling model definitions.} Configurations used for model-size experiments on the O'Doherty--Sabes dataset and the Allen Neuropixels dataset. Parameter counts exclude neuron and session embeddings.}
    \label{tab:param_scaling}
    \small
    \setlength{\tabcolsep}{6pt}
    \begin{tabular}{l c c c c}
        \toprule
        \textbf{Size} & \textbf{Hidden dim} & \textbf{Encoder depth} & \textbf{Decoder depth} & \textbf{Backbone params} \\
        \midrule
        Tiny   & 64  & 3 & 1 & 0.45M \\
        Small  & 96  & 4 & 2 & 1.30M \\
        Base   & 128 & 6 & 2 & 4.52M \\
        Large  & 192 & 8 & 3 & 8.33M \\
        XLarge & 256 & 12 & 4 & 20.00M \\
        \bottomrule
    \end{tabular}
\end{table}
\FloatBarrier

\clearpage
\section*{NeurIPS Paper Checklist}

\begin{enumerate}

\item {\bf Claims}
    \item[] Question: Do the main claims made in the abstract and introduction accurately reflect the paper's contributions and scope?
    \item[] Answer: \answerYes{}
    \item[] Justification: The main claims in the abstract and introduction are supported by the method and experimental results in Sections~\ref{sec:method} and~\ref{sec:results}.

\item {\bf Limitations}
    \item[] Question: Does the paper discuss the limitations of the work performed by the authors?
    \item[] Answer: \answerYes{}
    \item[] Justification: Section~\ref{sec:discussion} discusses the scope of the current evaluation and directions for future work.

\item {\bf Theory assumptions and proofs}
    \item[] Question: For each theoretical result, does the paper provide the full set of assumptions and a complete (and correct) proof?
    \item[] Answer: \answerNA{}
    \item[] Justification: The paper does not present theoretical results requiring proofs.

    \item {\bf Experimental result reproducibility}
    \item[] Question: Does the paper fully disclose all the information needed to reproduce the main experimental results of the paper to the extent that it affects the main claims and/or conclusions of the paper (regardless of whether the code and data are provided or not)?
    \item[] Answer: \answerYes{}
    \item[] Justification: Sections~\ref{sec:method}--\ref{sec:experiments} and Appendices~\ref{app:data_splits}--\ref{app:training} describe the model, data splits, baseline adaptations, training configuration, and evaluation protocol.

\item {\bf Open access to data and code}
    \item[] Question: Does the paper provide open access to the data and code, with sufficient instructions to faithfully reproduce the main experimental results, as described in supplemental material?
    \item[] Answer: \answerYes{}
    \item[] Justification: The neural recording datasets used in this work are publicly available from their original sources. The \ours{} source code is publicly available at \url{https://github.com/Jp-17/NeuroHorizon}.

\item {\bf Experimental setting/details}
    \item[] Question: Does the paper specify all the training and test details (e.g., data splits, hyperparameters, how they were chosen, type of optimizer) necessary to understand the results?
    \item[] Answer: \answerYes{}
    \item[] Justification: Section~\ref{sec:experiments} describes the experimental settings, with dataset splits in Appendix~\ref{app:data_splits} and training hyperparameters for the Perich--Miller experiments in Appendix~\ref{app:training}.

\item {\bf Experiment statistical significance}
    \item[] Question: Does the paper report error bars suitably and correctly defined or other appropriate information about the statistical significance of the experiments?
    \item[] Answer: \answerYes{}
    \item[] Justification: The main comparison, ablation, scaling, and transfer results report session means with standard errors of the mean (SEM) across sessions.

\item {\bf Experiments compute resources}
    \item[] Question: For each experiment, does the paper provide sufficient information on the computer resources (type of compute workers, memory, time of execution) needed to reproduce the experiments?
    \item[] Answer: \answerYes{}
    \item[] Justification: Appendix~\ref{app:training} reports the hardware, GPU memory usage, and training and evaluation times for the main experiments.

\item {\bf Code of ethics}
    \item[] Question: Does the research conducted in the paper conform, in every respect, with the NeurIPS Code of Ethics \url{https://neurips.cc/public/EthicsGuidelines}?
    \item[] Answer: \answerYes{}
    \item[] Justification: We have reviewed the NeurIPS Code of Ethics and confirm that our research conforms to it.

\item {\bf Broader impacts}
    \item[] Question: Does the paper discuss both potential positive societal impacts and negative societal impacts of the work performed?
    \item[] Answer: \answerNA{}
    \item[] Justification: This work is foundational research on neural population forecasting for scientific study. It is not tied to societal applications or deployments, and we do not foresee direct societal impacts from the work presented.

\item {\bf Safeguards}
    \item[] Question: Does the paper describe safeguards that have been put in place for responsible release of data or models that have a high risk for misuse (e.g., pre-trained language models, image generators, or scraped datasets)?
    \item[] Answer: \answerNA{}
    \item[] Justification: The released asset is research code for offline neural forecasting and does not include sensitive data or high-risk pretrained models.

\item {\bf Licenses for existing assets}
    \item[] Question: Are the creators or original owners of assets (e.g., code, data, models), used in the paper, properly credited and are the license and terms of use explicitly mentioned and properly respected?
    \item[] Answer: \answerYes{}
    \item[] Justification: The datasets, software, and baseline models used in this work are appropriately credited through citations to their original sources.

\item {\bf New assets}
    \item[] Question: Are new assets introduced in the paper well documented and is the documentation provided alongside the assets?
    \item[] Answer: \answerYes{}
    \item[] Justification: The new \ours{} code repository organizes the model implementation, configurations, and training and evaluation scripts with accompanying installation and usage documentation.

\item {\bf Crowdsourcing and research with human subjects}
    \item[] Question: For crowdsourcing experiments and research with human subjects, does the paper include the full text of instructions given to participants and screenshots, if applicable, as well as details about compensation (if any)?
    \item[] Answer: \answerNA{}
    \item[] Justification: This research does not involve crowdsourcing or human-subject research.

\item {\bf Institutional review board (IRB) approvals or equivalent for research with human subjects}
    \item[] Question: Does the paper describe potential risks incurred by study participants, whether such risks were disclosed to the subjects, and whether Institutional Review Board (IRB) approvals (or an equivalent approval/review based on the requirements of your country or institution) were obtained?
    \item[] Answer: \answerNA{}
    \item[] Justification: This research does not involve human subjects.

\item {\bf Declaration of LLM usage}
    \item[] Question: Does the paper describe the usage of LLMs if it is an important, original, or non-standard component of the core methods in this research? Note that if the LLM is used only for writing, editing, or formatting purposes and does \emph{not} impact the core methodology, scientific rigor, or originality of the research, declaration is not required.
    \item[] Answer: \answerNA{}
    \item[] Justification: LLMs were not involved in the core method development of this research.

\end{enumerate}

\end{document}